\documentclass[runningheads]{llncs}

\usepackage[utf8]{inputenc}
\usepackage{graphicx}
\usepackage{dirtytalk}
\usepackage{multirow}
\usepackage{tikz}
\usepackage{tabularx}
\usepackage{adjustbox}
\usepackage{enumitem}
\usepackage{makecell}
\usepackage{subcaption}
\usepackage{caption} 
\usepackage{hyperref}
\usepackage{tabu}
\usepackage{caption}

\newcommand{\itelos}{\textit{iTelos}}

\title{\itelos\  - Purpose Driven Knowledge Graph Generation}

\author{Fausto Giunchiglia\orcidID{0000-0002-5903-6150}\inst{1} \and
Simone Bocca\orcidID{0000-0002-5951-4589}\inst{1} \and Mattia Fumagalli\orcidID{0000-0003-3385-4769}\inst{2} \and Mayukh Bagchi \orcidID{0000-0002-2946-5018}\inst{1} \and Alessio Zamboni\orcidID{0000-0002-4435-1748}\inst{1}}

\institute{Department of Information Engineering and Computer Science (DISI),\\ University of Trento, Italy\\
\email{\{fausto.giunchiglia,simone.bocca,mayukh.bagchi,alessio.zamboni\}@unitn.it}
\and
Conceptual and Cognitive Modeling Research Group (CORE),\\
Free University of Bozen-Bolzano, Bolzano, Italy\\
\email{\{mattia.fumagalli@unibz.it\}}
}

\date{March 2021}

\begin{document}

\maketitle

\begin{abstract}
\vspace{-0.2cm}

When building a new application we are more and more confronted with the need of reusing and integrating pre-existing knowledge, e.g., ontologies, schemas, data of any kind, from multiple sources. Nevertheless, it is a fact that this prior knowledge is virtually impossible to reuse \textit{as-is}. This difficulty is the cause of high costs, with the further drawback that the resulting application will again be hardly reusable. It is a negative loop which consistently reinforces itself. \itelos\footnote{\itelos\ shares with \textit{Telos}\cite{mylopoulos1990telos} the focus of the \textit{purpose} (as from the name), with the latter being a language focused on how to represent it, the former being a methodology focused on how to exploit it in order to build suitable applications.} is a general purpose methodology aiming at minimizing as much as possible the effects of this loop. 
\itelos\ is based on the intuition that the \textit{data level} and the \textit{schema level} of an application should be developed independently,  thus allowing for maximum flexibility in the reuse of the prior knowledge, but under the overall guidance of  the needs to be satisfied, formalized as \textit{competence queries}. This intuition is implemented by codifying all the requirements, including those concerning \textit{reuse}, as part of an a-priori defined \textit{purpose}, which is then used to drive a \textit{middle-out} development process where the application schema and data are continuously aligned.
\vspace{-0.2cm}

\end{abstract}


\vspace{-0.1cm}
\section{Introduction} \label{sec1_introduction}

Once upon a time, one would design an application \textit{top-down} starting from the requirements down to implementation, without thinking of the data: they would be generated by the system in production with, at most, the need of initializing them with data from the legacy systems being substituted. Nowadays, more and more, we are designing systems which, at the beginning but also when in production, must be integrated with data coming from other systems, possibly from third parties. Some examples are the health systems which integrate personal data coming from multiple institutions and B2C applications exploiting the big data available on the Web, e.g., open data, or streaming data. 

The key aspect of this reuse problem is how to handle the \textit{semantic heterogeneity} which arises any time there is the need to perform data integration across multiple sources. This problem has been extensively studied in the past and two main approaches have been proposed. 
The first is using \textit{ontologies} to agree on a fixed language or schema to be shared across applications \cite{2017-obdi-for-mdee}. 
The second is the use of \textit{Knowledge Graphs (KGs)} and the exploitation of the intrinsic  flexibility and extensibility they provide \cite{2019-Kejriwal-KG}, as the means for facilitating the adaptation and integration of pre-existing heterogeneous data.
However the problem is far from being solved. When developing an application, no matter whether one exploits ontologies or KGs, it is impossible to reuse the pre-existing knowledge \textit{as-is}. There is always some specificity which makes the current problem in need of dedicated development, with the further drawback that the resulting application is, again, hardly reusable. 

 In this paper, we propose \itelos, a general purpose methodology 
 whose main goal is to minimize as much as possible the negative effects and the high costs deriving from this loop.
 \itelos\ exploits all the previous results, in particular, it is crucially based on the use of ontologies and KGs. At the same time, \itelos\ takes a step ahead by providing a precise specification of the process by which an application should be developed, focusing on how to effectively reuse data from multiple sources. \itelos\ is based on three key assumptions:
 
 \begin{itemize}
 \vspace{-0.1cm}
 \item the \textit{data level} and the \textit{schema level} of an application should be developed independently,  thus allowing for maximum flexibility in the reuse of the prior data and schemas, e.g., ontologies, but under the overall guidance of  the needs to be satisfied, formalized as \textit{competence queries};
 \item Data and schemas to be reused, as well as competence queries, should be decided before starting the development, as precisely as possible, and defined a priori as part of an application \textit{purpose}. Additionally, the purpose is assumed to specify a set of constraints specifying how much the satisfaction of each of its three elements is allowed to influence the satisfaction of the other two;
 \item the purpose should be used to drive a \textit{middle-out} development process where the successive evolutions of the application schema and data, both modeled as KGs, are continuously aligned \textit{upwards} (bottom-up) with the reference schemas to be reused and \textit{downwards} (top-down) with the data to be reused.
 \vspace{-0.5cm}
 \end{itemize}
This paper is organized as follows. 
In Section \ref{sec2-purpose} we provide a description of the purpose structure. 
In Section \ref{sec4_process} we describe the \itelos\ middle-out process. In Section \ref{sec5_reuse}  we provide a detailed description of how \itelos\ enhances the reusability of the available data. In Section
\ref{sec6_share} we describe how \itelos\ enhances the sharability and future reusability of the application KG. In Section \ref{sec7_evaluation} we provide the main alignment metrics used to ensure that the \itelos\ middle-out process stays within the constraints specified by the purpose. Section \ref{sec8_case} describes a set of case studies by which \itelos\ has been progressively evaluated and refined. Finally, Section \ref{sec9_conclusion} closes the paper with the conclusions.

\begin{figure}[ht!]
    \vspace{-0.6cm}
    \centering
    \includegraphics[width=8cm,height=4cm]{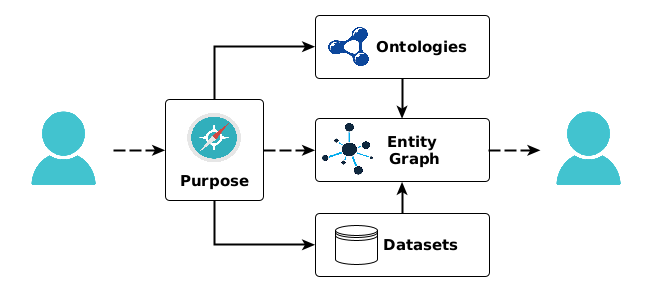}
    \vspace{-0.2cm}
    \caption{The \itelos\ approach.}
    \label{fig:approach}
    \vspace{-0.3cm}
\end{figure}

\section{The purpose}
\label{sec2-purpose}

A very high level view of the \itelos\ process is depicted in Figure \ref{fig:approach}. The \itelos\ logical sequence is represented by the dashed lines, where the \textit{User} provides an informal specification of the problem she wants to solve, the \textit{Purpose}, and receives in output a KG, called, in Figure \ref{fig:approach}, the \textit{Entity Graph}. The concrete factual process is represented by the four solid lines which indicate the reuse of the prior knowledge, represented at the data lavel as \textit{Datasets} and at the schema level as \textit{Ontologies}. 

The purpose contains four main elements, as follows:

\begin{itemize}
\vspace{-0.1cm}
    \item the functional requirements of the application, i.e., the needs to be satisfied, that we assume to be ultimately formalized as \textit{competence queries (CQ)} \cite{gruninger1995role}. 
    \item a set of datasets to be reused, as encoded in pre-existing applications, that we assume that, as part of the overall \itelos\ strategy, are progressively developed as KGs, thus facilitating their future reuse. 
    \item a set of pre-existing \textit{reference schemas}, ontologies but not only, whose reuse will facilitate the development of common schemas shared by future applications. It is important to notice that 
many such repositories are already available; for instance, LOV, LOV4IoT, and DATAHUB,\footnote{See, respectively: \url{https://lov.linkeddata.es/}, \url{http://lov4iot.appspot.com/}, \url{https://old.datahub.io/}.} are three among the most relevant repositories
On top of this, in order to facilitate the adoption of \itelos, a new repository, called \textit{LiveSchema},\footnote{See: \url{http://liveschema.eu/}.} is under construction, where reference schemas are annotated by a very rich set of metadata, see, e.g., \cite{dutta2015mod,2020-KR}, with the goal of automating as much as possible the \itelos\ process.
\item a set of metrics,  whose goal is to define the boundaries within which any of the three key elements of the purpose, \textit{viz.} competence queries, reference schemas, data to be reused, can be revised, in terms of use and reuse. 
\vspace{-0.1cm}
\end{itemize}
A crucial design decision in the structure of the purpose, which reflects into the overall \itelos\ process, as depicted in Figure \ref{fig:approach}, is that the data level and the schema level are kept distinct and independent, and that they are represented via two different types of KG. This assumption is key,  as it allows to split the problem of reusing existing data from the problem of facilitating the sharability and future reuse of the KG being developed.
Data level KGs, that we call \textit{Entity Graphs (EGs)}   are graphs where nodes are \textit{entities} (e.g., my cat \textit{Garfield}), decorated with data property values describing them, and where links are object properties which describe the relations holding between any two such entities. 
Schema level KGs, that we call \textit{entity type (etype) Graphs (ETGs)}, namely KGs which define the schema of EGs. In other words, for each EG there is a corresponding ETG which defines its schema. In ETGs nodes are \textit{etypes}, namely classes of entities (e.g., the class \textit{cat)}, each described by a set of data properties and by a set of object properties which define the range of links allowed among the nodes of the EG defined by the ETG. 

\textit{Datasets} and \textit{Ontologies} in Figure \ref{fig:approach} are, respectively examples of EGs and ETGs.
In \itelos\
EGs are used to represent: (i) the entity graph produced at the end of the process (see Figure \ref{fig:approach}) and, optionally, depending on their availability, (ii) the input datasets. ETGs and etypes are used to represent (i) the schema of the output EG, (ii) the reference schemas and ontologies and, also, ultimately, when fully formalized, (iii) Competency queries. This uniformity of representation provides two major advantages. The first is that, in practice, ETGs become syntactic sugar for a Description Logic (DL) TBOX while, dually, EGs become (syntactic sugar for) a (DL) ABOX, thus allowing for the usual DL reasoning. In particular, this allows us to see CQs as actual queries to the EG. The second is that, this uniformity of representation at the data and at the knowledge level can be used to implement, via an encoding of ETGs and EGs in Formal Concept Analysis (FCA) \cite{ganter2012formal}, various Machine Learning techniques which simplify the development of the output KG \cite{2020-KR}.

\begin{figure}[ht!]
\vspace{-0.6cm}
    \centering
    \makebox[\textwidth][c]{\includegraphics[width=15cm,height=3.8cm]{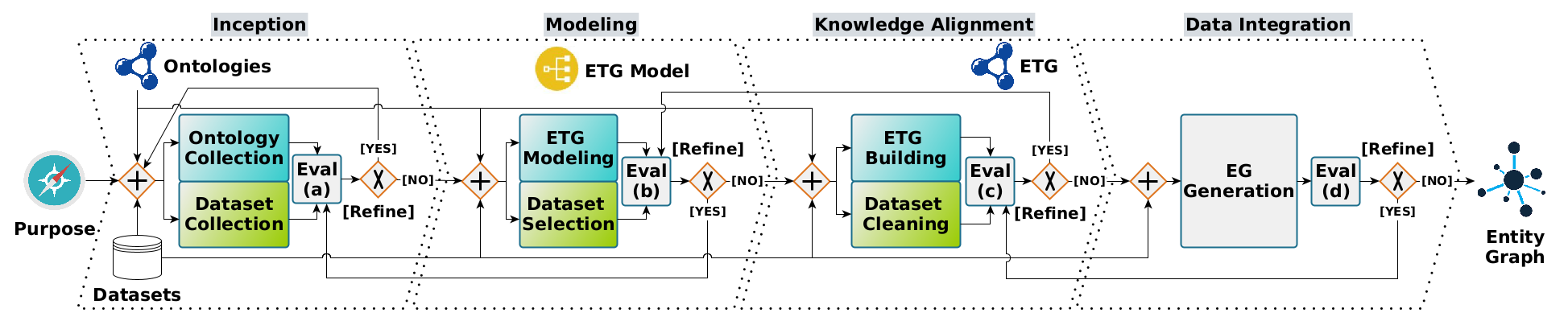}}
    \caption{The \itelos\ Process.}
    \label{fig:pipeline}
    \vspace{-0.9cm}
\end{figure}

\section{The Process} \label{sec4_process}

Figure \ref{fig:pipeline} instantiates the dashed line that in Figure \ref{fig:approach} produces the Entity Graph from the Purpose, into an overall process articulated in four phases. The purpose of each such phase can be synthetically described as follows:

\begin{itemize}
\vspace{-0.2cm}
    \item \textit{Inception}: takes in input the user's purpose and formalizes it, to the extent that it is needed, into a set of CQs and a first validated collection of input datasets and input ontologies;
    \item \textit{Modeling}: builds the ETG model by taking into account CQs and datasets;
    \item \textit{Knowledge Alignment}: builds a shareable ETG based on the model designed previously, via the reuse of the selected reference ontologies;
    \item \textit{Data Integration}: builds the output EG by integrating the input datasets into the ETG.
    \vspace{-0.1cm}
\end{itemize}
Let us analyze these phases in detail. 
The \textit{Inception} phase takes as input the purpose and collects the ontologies and datasets needed to build the target EG. The main goal is to finalize the functional requirements while enabling the reuse of existing resources. Because of this, the purpose is formalized, starting from its initial natural language description, into a list of CQs. 
CQs are then matched to the input datasets and ontologies thus implementing the activities of \textit{Ontology Collection} and \textit{Dataset Collection}, as from Figure \ref{fig:pipeline}. Depending on the process, the final list of collected datasets and ontologies may extend and also partially overlap with the input list.
The key observation is that matching CQs and (the schemas of the) datasets is crucial for the success of the project. A little coverage would mean that there are not enough data for the implementation of the EG and that there should be a revision of the CQs; 
whenever this is the case, the process we follow is inspired by the work in \cite{gruninger2019ontology}.

The \textit{Modeling} phase receives in input the ontologies and datasets previously collected, as well as the CQ list. The main objective of this phase is to build the most suitable model of the ETG to be used as the schema of the final EG, which in Figure \ref{fig:pipeline} is called the \textit{ETG model}. In practice, the ETG model includes all the etypes and properties needed to represent the information required by each CQ, possibly extended by extra etypes and properties suggested by the datasets. This extension is optional but suggested to allow for future expansions, given that the availability of data would make this step low cost, in particular, if taken into account since the early stages. 
The main objective of \textit{ETG Modeling} activity is to compose such etypes and properties into an EER model, essentially transforming the CQ list into an ETG. In parallel, the \textit{Dataset Selection} activity finalizes the selection of those datasets which are effectively required, pruning away useless data resources.

The \textit{Knowledge Alignment} phase takes as input the ETG model previously generated, plus the selected set of datasets and reference ontologies. The main objective of this phase is to enhance the sharability of the final EG, building in turn a shareable ETG that fits in the best possible way the datasets to be integrated. The input ETG model is itself a possible solution. However the set of reference ontologies provides more possibilities, in terms of etypes and properties available, that can be adopted to implement a final ETG (called ETG in Figure \ref{fig:pipeline}) easier to share and reuse. The \textit{ETG Alignment} activity is implemented via the Machine Learning algorithm described in \cite{2020-KR}. Such an algorithm takes as input the set of reference ontologies and the ETG model and it outputs the final version of the ETG. The resulting ETG is verified for compliance with the input datasets before the final approval.
In parallel the \textit{Dataset Cleaning} activity performs the final cleaning of the datasets consistently with the ETG. Here, the main objective is the alignment of the data types and formats of the datasets, with the etypes and properties of the ETG.

The last phase is \textit{Data Integration}. Its input consists of the ETG and the cleaned datasets. The objective is to build the EG integrating the schema and data resources, as adapted to the purpose in the previous phases. Unlike the previous ones, the data integration phase consists of a single activity, that we call \textit{EG Generation}, that merges together the schema and data layers creating the final EG.
 To do that, the ETG and datasets are provided in input to a specific data mapping tool, called \textit{KarmaLinker}, which consists of the \textit{Karma} data integration tool \cite{knoblock2015exploiting} extended to perform Natural Language Processing on short sentences (i.e., what we usually call \textit{the language of data}) \cite{2016-Bella1}. 
The first activity in this phase maps the data to the etypes and properties of the ETG. The following step is the generation of the entities that are then matched and, whenever they are discovered to be different representations of the same real world entity, merged.
These activities are fully supported by Karmalinker.
The above process is iteratively executed over the list of datasets selected in the previous phase, processed sequentially. The process concludes with the export of the EG into an RDF file.

The key observation is that the desired middle-out convergence, as described in the introduction, has been implemented in two separate sub-processes, executed in parallel within each phase, one operating and the schema level, the other at the data level (blue and green boxes in Figure \ref{fig:pipeline}).  During this process, the initial purpose keeps evolving building the bridge between CQs, datasets and ontologies. To enforce the convergence of this process, and also to avoid making costly mistakes, each phase ends with an evaluation activity (\textit{Eval} boxes in Figure \ref{fig:pipeline}).
The specifics of this activity are described in Section \ref{sec7_evaluation}. Here it is worth making two observations. The first is that this evaluation is driven by the non-functional requirements provided by the purpose. The second is that, within each phase, the evaluation aims to verify that the target of that phase is met, namely: aligning CQs with datasets and ontologies in phase 1, thus maximizing reusability; aligning the ETG model with the datasets in phase 2, thus guaranteeing the success of the project; and aligning ETG and ontologies in phase 3, thus maximizing sharability. The evaluation in phase 4 has the goal of checking that the final EG satisfies the requirements specified by the purpose. As from  Figure \ref{fig:pipeline}, a failure of the evaluation in any of the steps causes the process to go back to the evaluation step of the previous phase. In the extreme case of a major early design mistake it is possible to go back from phase 4 to phase 1. 
\vspace{-0.1cm}

\section{Enhancing data reuse} 
\label{sec5_reuse}

Reusability is enhanced during the phases of inception and modeling, whose main goal is to progressively transform the specifications from the purpose into the ETG Model. This process happens according to the following steps:

\begin{enumerate}
    \item generation of a list of natural language sentences, each informally defining a CQ, as implicitly or explicitly implied by the purpose; 
    \item generation of a list of relevant etypes and corresponding properties, which formalize the informal content of CQs, as from the previous step;
    \item selection of the datasets whose schema informally matches the CQs, as from the previous step;
     \item generation of a list of etypes with associated properties, from the selected datasets, which match the etypes and properties from the CQs;
     \item construction of the ETG model;
\end{enumerate}
 \noindent
 Steps 1-4 happen during inception, while step 5 happens during the modeling phase. Most of the work is definitely done during inception, while modeling just exploits and selects from the choices made during the previous phase with, possibly, the opportunity of backtracking in case the evaluation does not produce the desired results. 
 The key observation is that, in both phases, the three types of resources involved (i.e., CQs, ontologies, datasets) are handled through a series of three iterative executions, each corresponding to a specific category, following a decreasing level of reusability. The categories are defined as follows:

\vspace{0.2cm}
\noindent
\textit{Common}: this category involves resources associated with aspects that are common to all domains, also outside the domain of interest. Usually, these resources correspond to abstract etypes specified in \textit{upper level ontologies} \cite{DOLCE}, e.g., \textit{person}, \textit{organization}, \textit{event}, \textit{location}, and/or to etypes from very common domains, usually needed in most applications, e.g., \textit{Space} and \textit{Time}. These latter types of etypes and properties correspond to what in knowledge organization are called \textit{Common Isolates} \cite{srr67}. The data that are found in Open Data sites as well the ontologies which can be found in the repositories mentioned above are examples of common resources.
   
\vspace{0.2cm}
\noindent 
  \textit{Core}: this category involves resources associated with the more core aspects of the domain under consideration. They carry information about the most important aspects considered by the purpose, information without which it would be impossible to develop the EG. Consider for instance  the following purpose:
    
  \vspace{0.1cm}
  \noindent
        "\textit{There is a need to monitor Covid-19 data in the Trentino region (Italy), to understand the  diffusion  of  the  virus and  the  social  restriction  caused  by  the  virus,  with  the possibility to identify new outbreaks.}"\footnote{This example, as well as all the follow-up material, as described below in the paper, has been extracted from the project with title \textit{"Integration of medical data on Covid-19"} developed by Antony, N., Gotca, D., Jyate, M., Donini, L. The complete material and description of this project can be found at the URL \url{https://github.com/UNITN-KDI-2020/COVID-data-integration}. }

\vspace{0.1cm}
\noindent
    In this example, core resources could be those data values reporting the number of Covid-19 infections in the specified region. Examples of common resources are the data of certain domains, e.g., public sector facilities (e.g., hospitals, transportation, education),  domain specific ontologies that can be found again in the repositories above, as well as domain specific standards (e.g., Health, interoperability standards of various types). In general, data are harder to find than ontologies, in particular when they are about the private sector, where they carry economic value and are often collected by private bodies.

\vspace{0.2cm}
\noindent
\textit{Contextual}: this last category involves resources that carry specific, possibly unique, information from the domain of interest. These are the resources whose main goal is to create added value. If core resources are necessary for a meaningful application, contextual resources are the ones which can make the difference with respect to the competitors. In the above example, examples of contextual resources can be those data describing the type of social restrictions adopted to contrast the virus. At the schema level, contextual etypes and properties are those which differentiate the ontologies which, while covering the same domain, actually present major differences. \cite{2017-ICCM} presents a detailed quantitative analysis of how to compare these ontologies. Data level contextual resources are usually not trivial to find, given their specificity and intrinsic. In various applications we have developed in the past, this type of data have turned out to be a new set of resources those had to be generated on purpose for the application under development, in some cases while in production. 

The overall conclusive observation is that the availability of resources, and of data in particular, decreases from top to bottom. In parallel to reusability and sharability as well as, to a large extent, also to a decrease in cost, going from more infrastructural data to more application specific data. As performed during the evaluation, the consequent considerations on feasibility are among the drivers that guide the development of the EG. 

\begin{table}[]
\vspace{-0.9cm}
    \centering
    \caption{CQs categorized in the three reusability categories.}
    \vspace{-0.2cm}
    \begin{adjustbox}{max width=\textwidth}
    \begin{tabular}{|l|l|l|l|}
        \hline
        Number & Question & Action & Category \\
        \hline
        1 & How many cases in schools in Trentino? & Return the number of school cases in Trentino region & Contextual \\
        \hline
        2 & Are there schools that are closing? & Return whether schools are closed in Trentino region & Contextual \\
        \hline
        3 & Will the number of cases increase in Italy? & Return the infection prediction information in Italy & Core\\
        \hline
        4 & How many cases are in the RSA\footnotemark in Trentino region? & Return number of RSA cases in Trentino region & Contextual\\
        \hline
    \end{tabular}
    
    \end{adjustbox}
    \label{tab:cq}
    \vspace{-0.6cm}
\end{table}
\footnotetext{Residenza Sanitaria Assistenziale, Italian equivalent of extended care facility}
\noindent
 Let us see an example of how the six elaboration steps listed at the beginning of the section intermix with the three categories above.
 Table \ref{tab:cq} shows some CQs, extracted from the case study introduced above (step 1). Notice how, already in this step, CQs are categorized as being common, core or contextual (last column) and how this is done after translating the text from the purpose (left column) into something much closer to a requirement for the EG (central column). Table  \ref{tab:cq-etypes} then reports the etypes and properties extracted from the CQs (step 2). While this is not reported in the table for lack of space, these etypes and properties inherit the category from the CQs. Notice that it is possible to have an etype that is core or common or contextual with properties in all three categories; this being a consequence of the fact that an etype can be mentioned in multiple (types of) CQs.
 
 \begin{table}[]
\vspace{-0.9cm}
    \centering
    \caption{Etypes and properties from the CQs.}
    \vspace{-0.2cm}
    \begin{adjustbox}{max width=\textwidth}
    \begin{tabular}{|c|l|l|}
        \hline
        CQ Number & Etypes & Properties \\
        \hline
        1 & Covid\_status, Location & \makecell[l]{date, total\_number\_of\_cases, number\_of\_active\_cases,\\ number\_of\_new\_positive\_cases, number\_of\_deaths, number\_of\_recovered\_cases}\\
        \hline
        2 & Restriction, Location & \makecell[l]{location\_type, restriction\_type, closure\_start, closure\_end}\\
        \hline
        3 & Case\_projections, Case\_information, Location & \makecell[l]{location\_type,  date, mean\_of\_est.infections,\\ lower\_bound\_of\_est.infections, upper\_bound\_of\_est.infections}\\
        \hline
        4 & RSA\_cases & \makecell[l]{date, number\_of\_RSA\_case, number\_of\_home\_care\_cases}\\
        \hline
    \end{tabular}
    \end{adjustbox}
    \label{tab:cq-etypes}
    \vspace{-0.6cm}
\end{table}

The next step is to select the datasets (step 3). Let us assume that the dataset used to (partially) answer the queries in Table \ref{tab:cq} generates the properties  matching the CQs as in 
 Table \ref{tab:data} (step 4). As an example of matching, compare the attribute  \textit{cases} in  Table \ref{tab:data} with the property \textit{total\_number\_of\_cases} of CQ n°1 in Table \ref{tab:cq-etypes}.
 Notice how the ordering of the analysis (from common to core to contextual) creates dependencies that may drive the choice of one dataset over another. As an example,  \textit{School} might be needed as a core etype, but to properly define it, we also need the common etype \textit{Location}. 
The final ETG, as constructed, during steps 5, is reported in Figure \ref{fig:eer}, containing etypes and properties from the tables above. 

 \begin{table}[ht!]
    \centering
    \vspace{-0.9cm}
    \caption{Dataset's attributes classified according to the reusability categories.}
    \vspace{-0.2cm}
    \begin{adjustbox}{max width=\textwidth}
    \begin{tabular}{|c|c|c|c|}
        \hline
        Attributes & Description & Type & Category   \\
        \hline
        cases & Number of new cases on the date of the report & int & Core\\
        \hline
        deaths & Number of deaths on the date of the report & int & Core\\
        \hline
        countriesAndTerritories & Country and territory name related to the records & string & Contextual\\
        \hline
        year & Year from the date of report& int& Common\\
        \hline
        month & Month from the date of report& int& Common \\
        \hline
        day & Day from the date of report& int & Common\\
        \hline
    \end{tabular}
    \end{adjustbox}
    \label{tab:data}
    \vspace{-0.5cm}
\end{table}

\section{Enhancing data sharability} 
\label{sec6_share}

The knowledge alignment phase aims to enhance sharability, by aligning and possibly modifying the ETG model to take into account the etypes and properties coming from the reference ontologies. The key observation is that the alignment mainly concerns the common and, possibly, the core types with much smaller expectations on contextual etypes. This is where the repositories listed in Section \ref{sec2-purpose} can make the difference. Notice that, in retrospect, the alignment with the most suitable ontology can enable the reuse of data, at least for what concerns common and sometimes core etypes. As an example, the selection of Google GTFS or FOAF as reference ontologies ensures the availability of a huge amount of data, a lot of which are open data. This type of decision should be made during inception; if discovered here, it might generate backtracking. 

We align the ETG model with the reference ontologies, by adapting the \textit{Entity Type Recognition (ETR)} process proposed in \cite{2020-KR}. This process happens in three steps as follows:

\vspace{0.1cm}
\noindent
\textit{Step 1: ontologies selection}. This step aims at selecting the set of reference ontologies that best fit the ETG model. As from \cite{2020-KR}, this selection step occurs by measuring each reference ontology according to two metrics, which allow:
    \begin{itemize}
    \vspace{-0.2cm}
        \item to identify how many etypes of the reference ontologies are in common with those defined in the ETG model,  and
        \item to measure a property sharability value for each ontology etype, indicating how many properties are shared with the ETG model etypes. 
    \end{itemize}
\noindent The output of this first step is a set of selected ontologies, which best cover the ETG model, and that have been verified fitting the dataset's schema, at both etypes and etype properties levels.

\begin{figure}[ht!]
\vspace{-0.6cm}
    \centering
    \includegraphics[width=11cm,height=3cm]{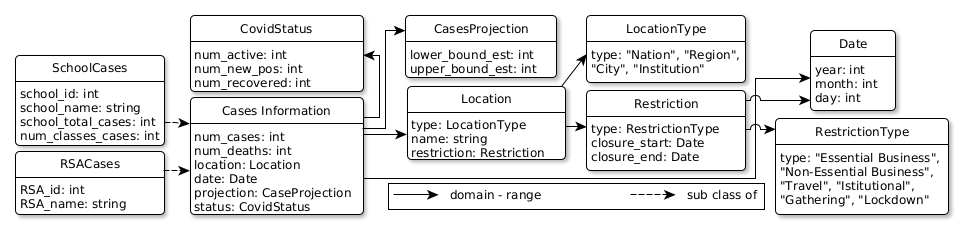}
    \vspace{-0.2cm}
    \caption{The example ETG.}
    \label{fig:eer}
    \vspace{-0.7cm}
\end{figure}

\noindent 
\textit{Step 2: Entity Type Recognition}(ETR). The main goal here is to predict, for each etype of the ETG model, which etype of the input ontologies, analyzed one at the time, best fits the ETG. In practice, the ETG model's etypes are used as labels of the classification task and, as mentioned in \cite{2020-KR}, the execution exploits techniques that are very similar to those used in ontology matching (see, e.g., \cite{2012-Giunchiglia3}). The final result is a vector of prediction values, returning a similarity score between the ETG model's etypes and the selected ontology etypes. 

\begin{figure}[ht!]
\vspace{-0.6cm}
    \centering
    \includegraphics[width=9cm,height=3.5cm]{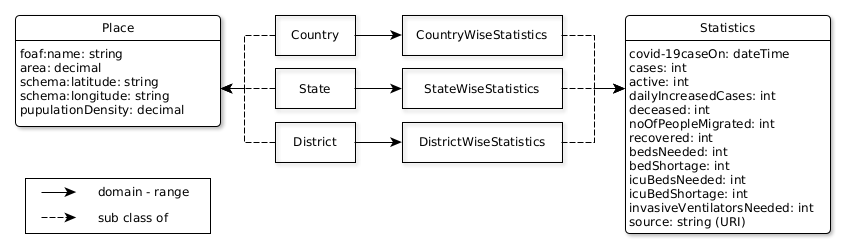}
    \vspace{-0.2cm}
    \caption{Portion of the \textit{Codo} reference ontology.}
    \label{fig:codo}
    \vspace{-0.7cm}
\end{figure}

\noindent
\textit{Step 3: ETG generation}. This step identifies, by using the prediction vector produced in the previous step, those etypes and properties from the ontologies which will be added to the final version of the ETG. Notice how this must be done while preserving the mapping with the datasets' schemas; whenever this is problematic, this becomes a possible source of backtracking. 
The distinction among common, core and contextual etypes and properties plays an important role in this phase and can be articulated as follows:
\begin{itemize}
\vspace{-0.2cm}
    \item The common etypes should be adopted from the reference ontology, in percentage as close as possible to 100\%. This usually results in an enrichment of the top level of the ETG model by adding those top level etypes (e.g., \textit{thing}, \textit{product}, \textit{event}, \textit{location}) that usually no  developer considers, because too abstract, but which are fundamental for building a highly shareable ETG where all properties are positioned in the right place. This also allows for an alignment of those \textit{common isolates} (see Section \ref{sec5_reuse}) for which usually a lot of (open) data are publicly available (e.g., \textit{street});
    \item The core etypes are tentatively treated in the same way as common etypes, but the results highly depend on the ontologies available. Think for instance of the GTFS example above;
    \item Contextual etypes and, in particular, contextual properties are mainly used to select among ontologies, the reason being that, the contextual properties allow distinguishing the most suitable among a set of ontologies about the same domain \cite{2020-KR}.
    \vspace{-0.1cm}
\end{itemize}
As an example, compare the ETG as from
 Figure \ref{fig:eer}, with the portion of the reference ontology \textit{Codo} \cite{codo2020} in Figure \ref{fig:codo}.\footnote{
 In this project, the matching between \textit{CODO} and the ETG model has been done manually.}  As it can be noticed, the two \textit{Codo} etypes \textit{Place} and \textit{Statistics} can be matched to the ETG etypes \textit{Location} and \textit{CasesInformation} via the ETG's etype  \textit{LocationType}. In this project the ETG model in Figure \ref{fig:eer} was adopted as the final ETG, as the authors preferred their original terminology and organization, over that of the selected reference ontology, which was allowed by the purpose they were provided.

\section{Alignment Evaluation} 
\label{sec7_evaluation}

\vspace{-0.1cm}
The evaluation is based on a set of metrics that are applied to CQs, ETGs, datasets, and dataset schemas, where the latter - when they are not KGs - are treated as sets of properties associated with one or more selected etypes of the ETG. These metrics are applied in the four phases of the \itelos\ process. For lack of space, below we describe only the application of three metrics to a specific phase.

Let $\alpha$ and $\beta$ be two generic \textit{element} sets, where an element can be an etype or a property. We have the following:

\vspace{0.1cm}
\noindent
 \textit{Coverage}. This metric allows to ascertain the extent to which an element set is appropriate for the purpose, and it is computed as the ratio between the intersection of $\alpha$ and $\beta$ and the whole $\alpha$. 
 \vspace{-0.2cm}
    \begin{equation}
    \label{eq1}
    Cov = (\alpha \cap \beta) / \alpha
    \end{equation}
    
\vspace{-0.1cm}
\noindent For instance, during the inception phase (\textit{Eval(a)} in Figure \ref{fig:pipeline}), \(Cov\) plays a central role in evaluating the \emph{reusability} of potential datasets (via dataset schemas) with respect to the CQs. For each dataset, a high value of \(Cov\), both applied to etypes and properties, implies that the dataset is highly appropriate for the purpose. A low value of \(Cov\) implies minimal overlap between the purpose and the dataset, the consequence being non-consideration of the dataset for reuse, and possible modification of the (underspecified) CQs. 
    
\vspace{0.1cm}
\noindent
\textit{Extensiveness}. This metric quantifies the proportional amount of knowledge provided by any element set (such as $\beta$), in terms of set of etypes or of sets of properties, with respect to the entire knowledge considered (here $\alpha$ and $\beta$)
 \vspace{-0.1cm}
    \begin{equation}
    \label{eq2}
    Ext = (\beta - (\alpha \cap \beta)) / ((\alpha + \beta) - (\alpha \cap \beta))
    \end{equation}
\noindent During the Modeling phase (\textit{Eval(b)} in Figure \ref{fig:pipeline}), the evaluation utilizes \(Ext\) to ensure that the ETG model maximally extends the set of CQs, which makes the model easily amenable for further evolution. To that end, a high value of \(Ext\) is evaluative of the fact that the ETG extends the scope of the CQs, by indicating a limited contribution of the CQs in generating the ETG model. On the other hand, low values of \(Ext\) are evaluative of the fact that CQs have contributed significantly towards enhancing the ETG model. 

\vspace{0.1cm}
\noindent
 \textit{Sparsity}. This metric quantifies the element-level difference between any number of similar  element sets, and is defined as the sum of the percentage of $\alpha$ not in $\beta$, and vice versa.
  \vspace{-0.2cm}
    \begin{equation}
    \label{eq3}
    Spr = ((\alpha + \beta) - 2(\alpha \cap \beta)) / ((\alpha + \beta) - (\alpha \cap \beta))
    \end{equation}
     
\vspace{-0.2cm}
\noindent 
In the knowledge alignment phase (\textit{Eval(c)} in Figure \ref{fig:pipeline}), our principal focus is to utilize \(Spr\) for ensuring the \emph{sharability} of the ETG. We incrementally enforce sharability by ensuring a required threshold of \(Spr\) between the ETG and each of the reference ontology, which indicates that the ETG contains axioms reflective of \emph{contextual knowledge}. 

\vspace{0.2cm}
\noindent
All the above metrics operate at the schema level. But they do not say anything about the results of integrating the datasets, as caused by the semantic heterogeneity existing among them. The situation is as follows. Assume that $D1$ and $D2$ are two datasets. Let us assume that, using \textit{KarmaLinker}, we analyze first $D1$ and then $D2$. Then we have the following possible situations:
\begin{itemize}
\vspace{-0.1cm}
    \item $D2$ contains an etype which is also contained by $D1$. In this case the EG is not extended; only the number of links across nodes is increased. We have two situations depending on whether an entity populates both datasets or only one. In the first case the entity will be enriched with a new set of property values with the possibility of contradictory information. In the second case, the two different sets of entities will values only for the properties  shared between $D1$ and $D2$. This case will result in an EG with a lot of missing links.
    \item $D2$ contains a new etype with respect $D1$. In this case, the EG is extended with new nodes and new links. We have again two situations depending in whether an entity populates both datasets or only one. In the first case the EG is nicely extended by attaching new links to the entities in $D1$. In the second case the two parts of the EG resulting from the import of the two datasets will be not connected.
\end{itemize}

\vspace{-0.1cm}
\noindent
The data driven metrics briefly introduced above are crucial for the evaluation of the quality of the final EG. At the moment, these characteristics of the EG have been evaluated with a set of techniques \textit{ad hoc}. A set of metrics for the evaluation of the quality of the final EG are under definition.

\section{Case studies} 
\label{sec8_case}

We have extensively evaluated and revised \itelos\ during the past three years (2018, 2019, 2020) as part of the Knowledge and Data Integration (KDI) class, a six credit course of the Master Degree in Computer Science of the University of Trento.\footnote{See  \url{https://unitn-kdi-2020.github.io/unitn-kdi-2020/} for  more details. This site contains the material used during the 2020 edition of the course and it consists of theoretical and practical lectures, as well as demos of the tools to be used, some of which have been mentioned above.} Table \ref{table-kdi} reports the information regarding the population involved and a number of projects. During this class, 2-5 students per group, must generate an EG using the pipeline above starting from a high level problem specification. Part of the task is also to identify the most suitable datasets and pre-process them. Datasets are usually found in \texttt{open data} repositories, but some are also scraped from the Web. The overall project has an elapsed time of 14 weeks during which students have to work intensely, even not full time. We estimate the overall effort each group puts into building an EG in around 4-8 person-months, depending on the case. 
As of today we have piloted 24 projects and 75 evaluations.

\begin{figure}[!htb]
\vspace{-0.6cm}
\minipage{0.52\textwidth}
  \includegraphics[width=\linewidth]{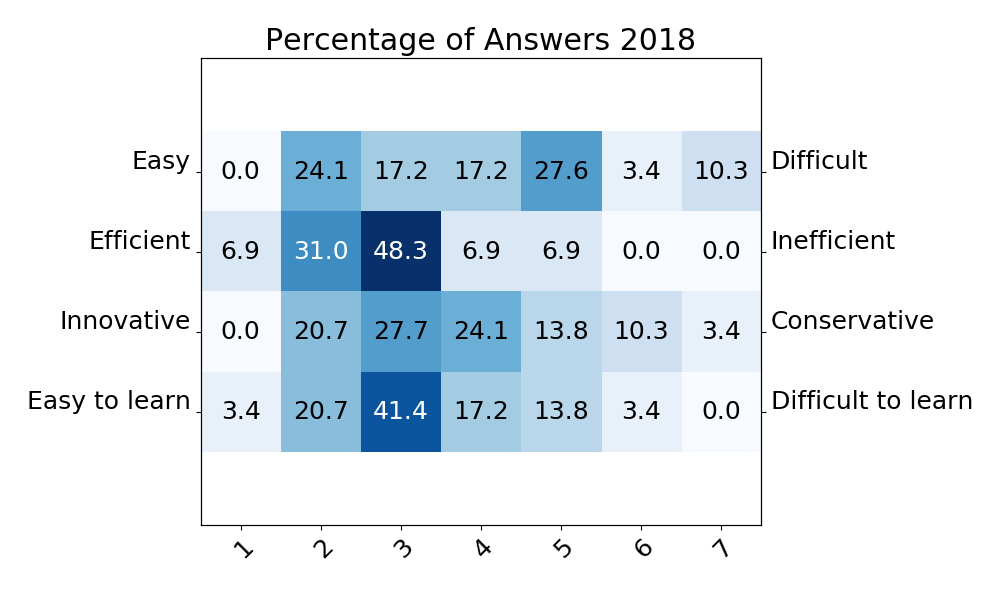}
\endminipage\hfill
\minipage{0.52\textwidth}
  \includegraphics[width=\linewidth]{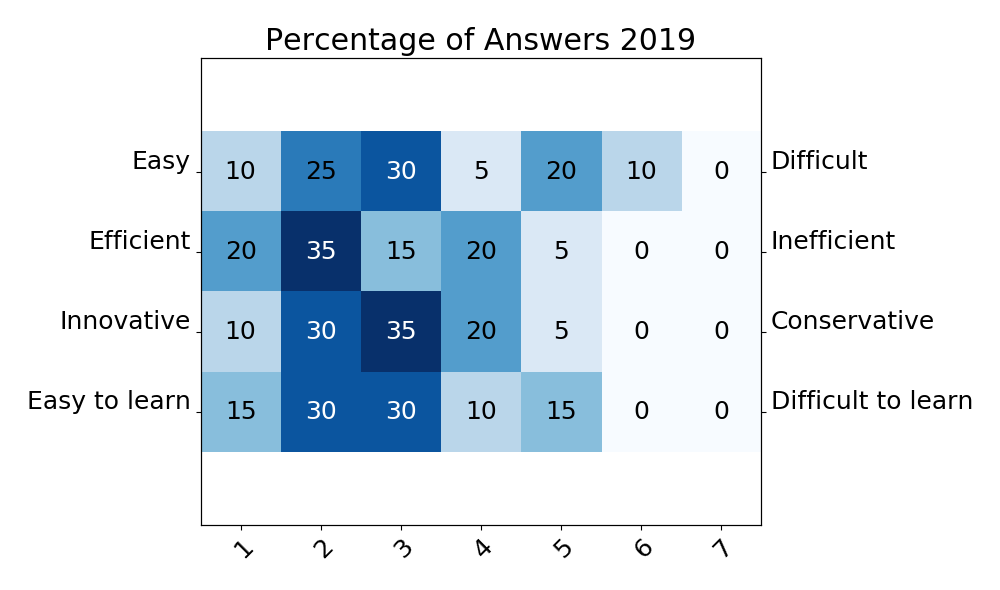}
\endminipage\hfill
\begin{center}
\minipage{0.52\textwidth}
\begin{center}
  \includegraphics[width=\linewidth]{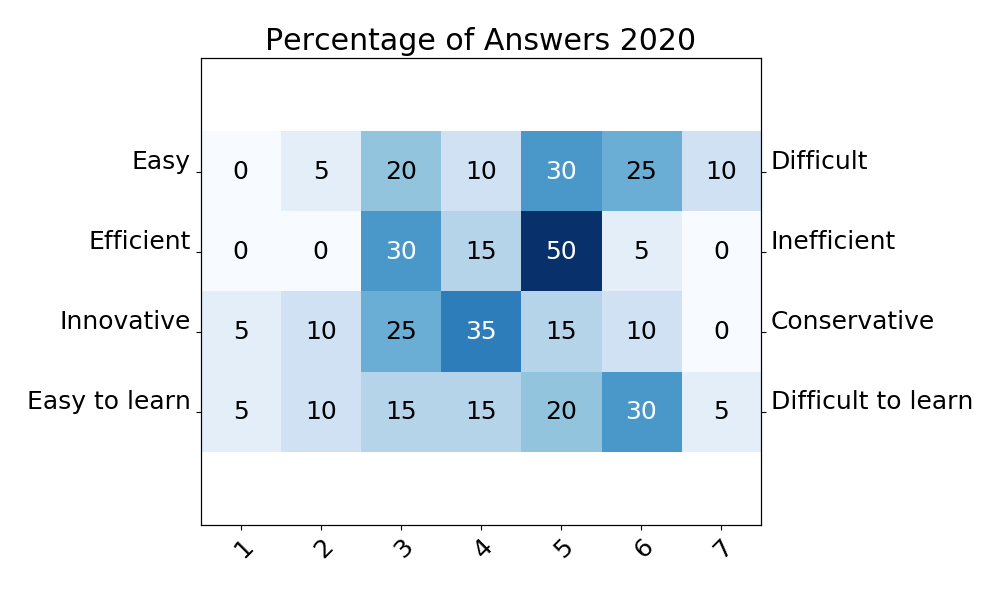}
\end{center}
\endminipage\hfill
\end{center}
\vspace{-0.2cm}
\caption{Usability of the methodology: (a) 2018, (b) 2019, (c) 2020}\label{fig:heatmap}
\vspace{-0.6cm}
\end{figure}

\setlength{\tabcolsep}{0.5em} 
{\renewcommand{\arraystretch}{1.2}
\begin{table}
\vspace{-0.9cm}
\caption{Evaluation's subjects in KDI class - 2018, 2019, 2020.}
\centering
\vspace{-0.2cm}
\begin{tabular}{|c|c|c|c|c|} 
\hline
        & \textbf{2018} & \textbf{2019} & \textbf{2020} & \textbf{Tot} \\
\hline
      \makecell[l]{\textbf{\# Students}} & 29 & 20 & 26 & 75\\
\hline
      \makecell[l]{\textbf{\# Project teams}} & 14 & 4 & 6 & 24\\
\hline    
      \makecell[l]{\textbf{\% Male}} & 69\% & 75\% & 95\% & 59\\
\hline
     \makecell[l]{\textbf{\% Female}} & 31\% & 25\% & 5\% & 16\\
\hline
     \makecell[l]{\textbf{\% Undergraduate}} & 20.7\% & 20\% & 25\% & 16.5\\
\hline
     \makecell[l]{\textbf{\% Post graduate}} & 79.3\% & 80\% & 75\% & 59\\
\hline
\end{tabular}
\vspace{-0.6cm}
\label{table-kdi}
\end{table}
}
\noindent
\noindent
The summary of the quantitative evaluation is reported in Figure~\ref{fig:heatmap}. We show three heat maps containing the answers to the seven quantitative questions submitted to users. All questions foresaw a scaled rating within the range $[1,7]$ where $1$ represents the most positive score and $7$ is the most negative. Our aim was to observe if, throughout the entire three-years period, the overall evaluation of the methodology revealed an increasing appreciation from the users. This point is validated by observing the heat maps related to the years 2018 and 2019. Indeed, the reader can appreciate the scores shifted to the left part of the heat map. Differently, from the evaluation collected in 2020, we observed a worsening of the scores (i.e. shift to the right). By combining the quantitative evaluation with the analysis of the qualitative evaluation, we concluded that during the first two years, we applied the \itelos\ methodology within an offline context where the face-to-face interaction was predominant. Instead in the last year, given the pandemic, we applied entirely the \itelos\ within an online context for the first time. This was an important test-bed since one of the long term objectives of \itelos\ is to support the distributed collaboration between people in creating and deploying high quality knowledge graphs. Hence, on the one hand, we observed how the refinement process of the methodology we setup for finding the best trade-off between effectiveness and efficiency in building knowledge graphs is proper. On the other hand, the deployment of the methodology within a complete online context triggered the need for further refinements for increasing its overall usability.

\noindent
\itelos\ users have also been asked to answer a questionnaire containing a set of qualitative queries to collect more fine-grained feedback. Below is a short description of the most important results, organized in \textit{strengths} and \textit{weaknesses}:

\begin{itemize}
\item \textit{(Strength)} the step by step, precisely articulated, \itelos\ process is easy to follow;
\item \textit{(Strength)} somewhat opposite to our expectations, the stepwise iterative evaluation process was perceived positively as a way for improving the overall outcome of the project as well as a support for the refinement of the Entity Graph; 
\item \textit{(Strength)} Two projects, carried out by PhD students, were also developed as part of EC projects, and required interaction with third party data providers. It turned out that the \itelos\ process 
adapts well to the cooperative specification among multiple parties, in particular in the case of data providers and KG developers;
\item \textit{(Weakness)} A wrong decision made in the early phases is quite difficult to remedy, with this possibility being very high during the inception phase;
\item \textit{(Weakness)} The work between the schema and the data layer is unbalanced in favour of the second, in particular during the \textit{informal modeling} phase. This complicates the synchronization of the work with the possibility of misalignments, mainly because of misunderstandings, that have to be handled very carefully by the project manager.
\end{itemize}
\section{Conclusions} 
\label{sec9_conclusion}

In this paper we have introduced \itelos, a novel methodology whose ultimate goal is to implement a \textit{circular} development process. By this we mean that the goal of \itelos\ is to enable the development of EGs via the  \textit{reuse} of already existing EGs and ETGs, while being simultaneously developed to be later easily \textit{reused} by other applications to come. 
%

\vspace{0.3cm}
\noindent
{\bf Acknowledgements} The research conducted by Fausto Giunchiglia, Mayukh Bagchi and Simone Bocca has received funding from the \emph{“DELPhi - DiscovEring Life Patterns”} project funded by the MIUR (PRIN) 2017. The research conducted by Alessio Zamboni was supported by the \emph{InteropEHRate} project, EC Horizon 2020 programme under grant number 826106.

\bibliographystyle{splncs04}
\bibliography{KnowDivePubs}

\end{document}